\documentclass[11pt]{amsart}
\usepackage{fullpage,prefix1,tikz}
\title{DGLA Actions: An Application in GR}
\author{Ryan Grady}
\address{Montana State University, Bozeman, MT, USA, ryan.grady1@montana.edu}

\DeclareMathOperator\bigw{\scalebox{.95}[1]{$\bigwedge$}}

\begin{document}

\begin{abstract}
This note serves two purposes: 1) define actions by differential graded Lie algebras, and 2) apply such differential graded Lie symmetry in general relativity (GR) to constrain the spacetime geometry on a neighborhood of infinity.
\end{abstract}

\maketitle
\thispagestyle{empty}

\section{Introduction}

Differential graded Lie algebras (dglas) are generalizations of Lie algebras which came into use nearly 100 years ago in the work of \'{E}lie Cartan and his moving frame formalism.  The objects of study are ``built up" from Lie algebras in a few simple steps: consider {\em graded} Lie algebras, i.e., the underlying vector space is graded and all structure/properties respect this grading; and secondly, turn on a differential making the underlying graded vector space a cochain complex. (Additionally, the differential is required to be a graded derivation of the Lie bracket.)

Just as Lie algebras encode infinitesimal symmetries of physical systems, dglas also are symmetry objects.  The additional structure of a dgla allows one to consider several symmetries which interact at once.  Differential graded Lie algebras are also one approach to formulating ``symmetry up to homotopy."  As we illustrate in the next section, dglas arise naturally when considering pure gauge theory.

In Section \ref{sect:2}, we define dglas and their actions carefully.  We also indicate how a dgla acts on a classical field theory. In Section \ref{sect:3}, we use dglas to enforce an algebra of Killing fields on a neighborhood of infinity for a spacetime. That is, we have a dgla which acts on (classical) Palatini--Cartan Theory.  In several examples, specifying the Killing algebra on a neighborhood of infinity significantly constrains the spacetime geometry.

In Section \ref{sect:2}, we define dglas and their actions.  We also indicate how a dgla acts on a classical field theory. In Section \ref{sect:3}, we use dgla actions to constrain spacetime geometry. That is, we have a dgla which acts on (classical) Palatini--Cartan Theory and determines the Killing algebra outside of a compact region.

A few clarifying comments are in order.  First, in Section \ref{sect:3} our symmetry algebra is supported on an honest neighborhood of infinity, i.e., on an open set which is the complement of a compact region.  As such, the method currently presented requires some extension/work to encode true asymptotic symmetries, e.g., BMS symmetry.  Approaching asymptotic symmetry through dgla/$L_\infty$ actions is work in progress with Filip Dul and Surya Raghavendran.  Secondly, symmetry in classical and quantum field theory naturally leads to conserved quantities and charges via Noether type theorems.  For instance, Wald and Iyer realized black hole entropy via Killing symmetry.  The dgla/$L_\infty$ approach to conserved quantities, e.g., entropy, in GR is also ongoing work with Dul, Raghavendran, and Changsun Choi, see \cite{Dul} for an introduction. A complimentary approach has also been presented by Blohmann \cite{Blo}.

\emph{This version of the note has an appendix which contains further remarks and details.  The appendix does not appear in the published proceedings version of this note.}

\subsection*{Acknowledgements}
I thank my collaborators on related projects Filip Dul,  Surya Raghavendran, and Changsun Choi. I am also appreciative of several discussions with Peter Hintz and Justin Corvino regarding the gluing of spacetimes. Finally, I thank the anonymous referee for several helpful suggestions.  The author is supported by the Simons Foundation under Travel Support/Collaboration 9966728.

\section{DGLAs and Actions}\label{sect:2}

The notion of smooth symmetry is typically encoded via Lie groups and their actions.  The infinitesimal symmetries---equivalently one parameter families of symmetries---are encoded by Lie algebras and their actions.  While both Lie groups and Lie algebras play an important role in mathematical physics, it is the action of Lie algebras which often appear in Lagrangian formulations of QFT.  In general relativity (GR), the Lie algebra of vector fields are the infinitesimal diffeomorphisms, while the Lie algebra of Killing fields consists of infinitesimal isometries.

Differential graded Lie algebras (dgla), to be defined shortly, are a richer mathematical structure which strictly generalize ordinary Lie algebras.  Perhaps the most familiar dgla is differential forms on spacetime, $M$,  valued in a Lie algebra $\mathfrak{g}$, $\Omega^\ast (M, \mathfrak{g})$.  This algebra plays a central role in pure gauge theory.  For a trivial bundle---there is a similar, but slightly more complicated version for non-trivial bundles---the connection one-form $A \in \Omega^1(M, \mathfrak{g})$ encodes how the gauge group $G$ is acting on tangent spaces (and their decomposition into vertical and horizontal subspaces) as we move around our spacetime. That is, the connection one-form can be thought of encoding a sort of local symmetry action.  A $\mathfrak{g}$-valued function, $\varphi \in \Omega^0 (M, \mathfrak{g})$, further encodes the gauge action on connections as one moves around the spacetime manifold. 

Recall that a graded Lie algebra is a graded vector space equipped with a degree zero (bilinear) bracket that is graded skew symmetric and satisfies the graded Jacobi relation.

\begin{definition}
Let $\mathbb{K}$ be a field.  A \emph{differential graded Lie algebra} over $\mathbb{K}$ consists of
\begin{enumerate}
\item A graded Lie algebra $(\mathfrak{g}^\bullet, [-,-])$ over $\mathbb{K}$, and
\item A $\mathbb{K}$-linear map of degree one $d \colon \mathfrak{g}^\bullet \to \mathfrak{g}^{\bullet +1}$.
\end{enumerate}
Such that 
\begin{itemize}
\item[(a)] The map $d$ is a differential, i.e., $d^2=0$, and
\item[(b)] The map $d$ is a derivation of the bracket, i.e., for all $X,Y \in \mathfrak{g}^\bullet$
\[
d[X,Y] = [dX,Y] + (-1)^{\lvert X \rvert} [X,dY],
\]
where $\lvert X \rvert$ denotes the degree of $X$.
\end{itemize}

\end{definition}

A (strict) map of dglas $\varphi \colon \mathfrak{g}^\bullet \to \mathfrak{h}^\bullet$ is a degree zero linear map which commutes with differentials and preserves brackets. General references for differential graded Lie algebras, their maps, and their actions (and the extension to $L_\infty$ algebras) are the book of Manetti \cite{Man} and Appendix A of the second volume of the tome by Costello and Gwilliam \cite{CG2}. 

Let us now consider actions of differential graded Lie algebras.  To begin, let $\mathfrak{g}$ be an ordinary Lie algebra.  A representation/module for $\mathfrak{g}$ is a vector space $V$ and a map of Lie algebras $\alpha \colon \mathfrak{g} \to \mathrm{End}(V)$, where $\mathrm{End}(V)$ is the Lie algebra consisting of linear self-maps of $V$ and is equipped with the commutator bracket. This data can be reorganized by equipping the direct sum $\mathfrak{g} \oplus V$ with the structure of a Lie algebra such that
\[
0 \to V \xrightarrow{\; \; \psi \; \;} \mathfrak{g} \oplus V \xrightarrow{\;\; \phi \; \;} \mathfrak{g} \to 0
\]
is a short exact sequence of Lie algebras.  Indeed, given an action map $\alpha$ as above, define the Lie bracket on $\mathfrak{g} \oplus V$ by
\[
[[(X,v),(Y,w)]] := ([X,Y]_\mathfrak{g}, \alpha(X)(w) + \alpha(Y)(v)).
\]
To go the other way, suppose we have a Lie bracket $[[-,-]]$ on $\mathfrak{g} \oplus V$, we define an action map $\alpha \colon \mathfrak{g} \to \mathrm{End}(V)$ as follows. Given $X \in \mathfrak{g}$ and $w \in V$, consider $[[(X,0),(0,w)]]$.  This element will map to $0 \in \mathfrak{g}$ under the map $\phi$, so by exactness corresponds to a unique element $\tilde{w} \in V$. Hence, define $\alpha (X) (w) = \tilde{w}$. 

While the preceding paragraph may have seemed convoluted, it leads to an immediate definition of an action by a dgla.  In fact, it allows for a definition of one dgla $\mathfrak{g}^\ast$ acting on another dgla $\mathfrak{h}^\ast$, which is what appears when considering symmetries in the BRST/BV formalism.

\begin{definition}\label{defn2}
Let $\mathfrak{g}^\bullet$ and $\mathfrak{h}^\bullet$ be differential graded Lie algebras. An \emph{action} of $\mathfrak{g}^\bullet$ on $\mathfrak{h}^\bullet$ is the structure of a differential graded Lie algebra on $\mathfrak{g}^\bullet \oplus \mathfrak{h}^\bullet$ such that there is a short exact sequence
\[
0 \to \mathfrak{h}^\bullet \xrightarrow{\; \; \psi \; \;} \mathfrak{g}^\bullet \oplus \mathfrak{h}^\bullet \xrightarrow{\;\; \phi \; \;} \mathfrak{g}^\bullet \to 0
\]
with $\psi$ and $\phi$ being maps of dglas.\footnote{Recall the grading convention for the direct sum of graded vector spaces: $(\mathfrak{g} \oplus \mathfrak{h})^i = \mathfrak{g}^i \oplus \mathfrak{h}^i$.}
\end{definition}


\begin{example}
Given a dgla $(\mathfrak{g}^\bullet, d , [-,-])$, it acts on itself (considered as a dgla) via the adjoint action.  Indeed, the dgla structure on $\mathfrak{g}^\bullet \oplus \mathfrak{g}^\bullet$ is given by
\[
d_\oplus (X,X') := (dX, dX') \quad \text{ and } \quad [[(X,X'),(Y,Y')]]_\oplus := ([X,Y], [X',Y'] + [X,Y'] + [X',Y]).
\]
There is a similarly defined co-adjoint action as well.

\end{example}


\subsection{Actions on Classical Field Theories}

Given a classical field theory $(\cE,S)$, a symmetry action by a Lie algebra should be by symplectic vector fields on the space of classical solutions.  If we are interested in realizing symmetries as observables, e.g., as in Noether's Theorem, then we should actually ask that our action factors through the Lie subalgebra of Hamiltonian vector fields.

The BV formalism provides a resolution of the (functions on) the Euler--Lagrange locus. This resolution is with respect to natural gauge symmetries, we may wish to consider additional symmetries and their associated conserved currents/charges.  In the BV formalism, there is a natural (shifted) symplectic structure and corresponding Poisson bracket.  The condition for our symmetry action now becomes that $\mathfrak{g}$ acts via symplectic vector fields which commute with the vector field $\{S,-\}$. 

One mathematical formulation of actions/symmetries is provided in Part 3 of Volume 2 of Costello--Gwilliam \cite{CG2}.  Moreover, Costello and Gwilliam reformulate actions in terms of equivariant actions, see Definition 12.2.12 of \cite{CG2}. This equivariant action is the BV incarnation of our Definition \ref{defn2} from above.  The rough idea underlying Costello--Gwilliam's approach is to write every theory as a family of (deformed) BF or Chern--Simons theories.  As such, any classical field theory is itself presented by a Lie theoretic object.  The further action of $\mathfrak{g}$ on our theory can be rephrased in terms of a certain Maurer--Cartan element in an auxiliary ``action dgla"; this method is standard and simply an extension of writing a map between Lie algebras in terms of their associated Chevalley--Eilenberg cochains.  This Mauer--Cartan element then determines an action functional on an extended space of fields: the equivariant action functional. This process is reversible, so the symmetry action and the equivariant action functional are equivalent data.

In our application below, we will simply describe the explicit equivariant action. Note that for completeness, we should verify the equivariant classical master equation in the BV formulation of our theory; we do not do this here, details are given in \cite{GChoi}.

\section{Application: Geometry on a Neighborhood of Infinity via an Equivariant Action}\label{sect:3}

In this section, and for the applications below, we will assume familiarity with some classic notions and results in GR; Wald's text \cite{Wald} or Hawking--Ellis \cite{HE} contain details, arguments, and further references.

Following \cite{CS17, CS19}, we recall the 4D Palatini--Cartan action for a fixed spacetime manifold $M$.  Indeed, 
\begin{itemize}
\item Let $\cV \to M$ be a Minkowski (reference) bundle, so each fiber is Minkowski space $(V, \eta)$.
\item Let $P \to M$ be the $SO(3,1)$ principal (frame) bundle associated to $\cV$.
\item Let $\omega \in \cA_P$ be a connection, with curvature 
\[
F_\omega \in \Omega^2 (M, \mathfrak{so}(3,1)) \cong \Omega^2 (M, \bigw^2 V).
\]
\item Let ${\bf e} \colon TM \to \cV$ be a bundle isomorphism covering the identity. That is, ${\bf e}$ is a \emph{tetrad/coframe/vierbein} with ${\bf e} \in \Omega^1_{\text{nd}} (M, \cV)$, where the ``\text{nd}" denotes the non-degeneracy of the tetrad.
\item Finally, let $\Tr \colon \bigw^4 V \to \RR$ be the trace map induced from an orthonormal basis for $V$.
\end{itemize}
Then, the \emph{Palatini--Cartan action} is given by
\[
S_{\text{PC}} ({\bf e},\omega) = \int_M \Tr \left [ \frac{1}{2} {\bf e} \wedge {\bf e} \wedge F_\omega + \frac{\Lambda}{24} {\bf e}\wedge {\bf e} \wedge {\bf e} \wedge {\bf e} \right ],
\]
for \emph{cosmological constant} $\Lambda$.

Note that the tetrad, ${\bf e} \colon TM \xrightarrow{\; \simeq \;} \cV$, determines a metric $g_{\bf e} := {\bf e}^\ast \eta$ by pulling back the fiberwise metric on the reference bundle $\cV$.  The Euler--Lagrange Equations (EOMs) for $S_{\text{PC}}$ are given by
\[
{\bf e} \wedge d_\omega {\bf e} =0 \quad \text{ and } {\bf e} \wedge F_\omega + \frac{\Lambda}{6} {\bf e} \wedge {\bf e} \wedge {\bf e} =0.
\]
As is explained in Section 6.1 of \cite{Rom}, the first equation shows that the covariant derivative $d_\omega$ (and hence $\omega$ itself) is determined by the tetrad.  There is a unique torsion free, metric connection with respect to the pullback metric $g_{\bf e}$, its Levi--Civita connection. It then follows that the second equation can be written in terms of the Riemann curvature tensor and upon contraction with the tetrad recovers Einstein's field equations.

\subsection{Enforcing a Killing Algebra}

In this section and what follows, for simplicity, we restrict to the case $M = \RR^4$. 

Let $\mathfrak{p} := \RR^{3,1} \rtimes \mathfrak{so}(3,1)$ be the Poincar\'{e} algebra.  In particular, we have the isomorphism
\[
\mathfrak{p} \cong \mathfrak{iso}(\RR^{3,1}) = \mathrm{Lie} (\mathbf{Iso}(\RR^{3,1})),
\]
i.e., $\mathfrak{p}$ is the Lie algebra of Killing vector fields for the Minkowski metric.

Let us fix $\mathfrak{g} \subseteq \mathfrak{p}$  a Lie subalgebra.

Fix two radii $R >r \gg 0$ and a smooth cut-off function $\Upsilon \colon \RR^4 \to \RR$ such that $\Upsilon$ vanishes inside the sphere (centered at the origin) of radius $r$ and such that $\Upsilon$ is the constant function 1 outside the sphere of radius $R$. Let $U \subseteq \RR^4$ be the complement of the closed ball of radius $r$, i.e., $U$ is the region of spacetime outside of the inner sphere. Note that the cut-off function induces a natural ``extension by zero" map $i_\Upsilon \colon \Omega^\ast (U, \mathfrak{g}) \hookrightarrow \Omega^\ast (\RR^4, \mathfrak{g})$; we will often abuse notation and identify a form with its image under this map.

Let $\alpha \otimes X \in \Omega^\ast (U, \mathfrak{g})$ and decompose $\alpha$ as $\alpha = \alpha \otimes (X_T + X_R)$, where
\[
\alpha \otimes X_T \in \Omega^\ast (U, \RR^{3,1}) \quad \text{ and } \quad \alpha \otimes X_R \in \Omega^\ast (U, \mathfrak{so}(3,1)).
\]
(Note there is a slight abuse of notation as $\mathfrak{g}$ is only a sub algebra of $\mathfrak{p}$, nonetheless the decomposition just described descends to a $\mathfrak{g}$-valued form.) Note that given a Lie algebra element $X_T$, there is a natural ``shift" action on any tetrad; we will denote this action by $X_T \cdot {\bf e}$.
 Now define an equivariant action functional
\[
S^\mathfrak{g} ({\bf e}, \omega, \alpha \otimes X) := S_{\text{PC}}({\bf e}, \omega) +\frac{1}{2} \int_M \Tr \left [(\alpha \otimes X) \cdot {\bf e} \wedge {\bf e} \right ],
\]
where
\[
(\alpha \otimes X) \cdot {\bf e} \wedge {\bf e} = (X_T \cdot {\bf e} ) \wedge (X_T \cdot {\bf e}) \wedge \alpha \otimes X_R \in \Omega^\ast (M, \bigw^4 V).
\]

Varying the equivariant action $S^\mathfrak{g}$ now produces an additional term in the second EOM.  Suppose $({\bf e}, \omega)$ was already a classical solution to $S_{\text{PC}}$, then the addition condition is
\[
X_T \cdot {\bf e} \wedge \alpha \otimes X_R =0.
\]
For fixed $X =X_T \oplus X_R \in \mathfrak{g}$, this must hold for arbitrary $\alpha \in \Omega^\ast (U)$.  This implies that on $U$, $L_X {\bf e} = 0$, and consequently on $U$
\[
L_X g_{\bf e} = L_X ({\bf e}^\ast \eta) = 0.
\]

\subsubsection{The Poincar\'{e} Algebra} Let $\mathfrak{g} = \mathfrak{p}$, so the entire Poincar\'{e} Lie algebra. In four dimensions, this is a maximal set of Killing fields ($n(n+1)/2 =10$ for $n=4$) and determines the metric up to a constant.  Hence, the spacetime metric on our infinite region $U$ is just a constant multiple of the Minkowski metric.

Now, as the (ADM and Bondi) mass/energy can be computed as conserved quantities on the sphere at infinity, if $L_X g =0$ on our infinite region $U$ for all $X \in \mathfrak{p}$, then our spacetime has the same energy/mass of flat space, zero.  Under reasonable energy assumptions, e.g., the \emph{dominant energy condition}, the Positive Energy Theorem of Schoen--Yau and Witten implies that our spacetime is globally Minkowski space.

\subsubsection{Spherical Symmetry} Consider now the Lie algebra of infinitesimal rotations
\[
\mathfrak{g} =  \mathfrak{so}(3),
\]
with generators $\partial_t, L_1, L_2, L_3$ and non-trivial brackets determined by
\[
[L_i ,L_j ] = \epsilon_{ijk} L_k .
\]

In the case of vanishing cosmological constant, $\Lambda=0$, then Birkhoff's Theorem implies that our spacetime on the infinite region $U$ is just Schwarzchild spacetime.  

Spacetimes beyond globally Schwarzchild are possible via {\em spacelike} and {\em characterisric gluing} of spacetimes.  The gluing of solutions to the Einstein Equations has a long and beautiful history which is recently overviewed by Corvino \cite{Cor}.  In our case, one can use Theorem 1.1 of \cite{CR22} to create a spacetime which is Minkowskian at small radii and is Schwarzchild at large radii.  Other relevant gluing constructions are given in \cite{KU23} and \cite{MOT23}.


%
%

Note, if we allowed more interesting topology of the underlying space, then we could also obtain AdS-Schwarz and dS-Schwarz type geometries in a neighborhood of infinity, see \cite{SW09}.

%

%

\appendix

\section{Fundamental Concepts/Results}

\subsection{Mass as a Conserved Quantity}

As discussed many places, e.g., \cite{Wald}, \cite{MTW}, mass is a subtle concept in general relativity. The are several definitions of mass; three notable ones are ADM, Bondi, and Komar.  Similarly, there is a notion of momentum and energy corresponding to each of these three notions.  When their domains of definition overlap, Komar and ADM masses are equivalent.  Bondi is similar to ADM, but does not always agree on the nose.

One important point is that for ADM, Bondi, and Komar masses, they can be computed as conserved quantities at infinity.  

Komar mass is defined for \emph{stationary} spacetimes, i.e., those which possess a timelike killing field.  The Komar energy, from which the mass can be defined, is the conserved quantity associated to this killing field and can be computed as an integral over a sphere at infinity.

ADM and Bondi mass are defined for \emph{asymptotically flat} spacetimes. Such spacetimes admit a conformal infinity which contains special regions known as \emph{null infinity} and \emph{spacelike infinity}.  The \emph{Living Review} article \cite{Frau} is a good introduction to these topics, as is Chru\'{s}ciel's text \cite{PC}. The ADM mass is computed as a surface integral over a sphere at spacelike infinity, while the Bondi mass is computed at null infinity.

\subsection{Positive Mass Theorem}

A recent review is provided in \cite{Huang}, but let me just recall the general statement of the theorem.

\begin{theorem}[Positive Mass Theorem]
Let $(M,g)$ be an asymptotically flat spacetime of dimension $n$ (which satisfies the dominant energy condition).  If $3 \le n \le 7$ or $M$ is spin, then $E \ge \lvert P \rvert$ where $(E,P)$ is the ADM energy-momentum vector of $(M,g)$.  Moreover, as the ADM mass is given by $m = \sqrt{E^2-\lvert P \rvert^2}$, under the hypotheses $m \ge 0$. Finally, if $n=3,4$ and $m=0$, then $(M,g)$ can be isometrically embedded into Minkowski space.

\end{theorem}

The contributors to this theorem include Choquet-Bruhat--Marsden, Schoen--Yau, Witten, Chru\'{s}ciel, Hawking, and others.  The theorem now applies also to Bondi mass and some asymptotically AdS spacetimes.

\subsection{Birkhoff and Israel's Theorems}

Birkoff's Theorem is another object which is really more of a collection of results of many people, including Jebsen two years before Birkhoff published his result.  See \cite{Schmidt} for an overview of the many related results.  The basic idea of Birkhoff's Theorem is that any spherically symmetric solution of the vacuum Einstein equations must be static and asymptotically flat.  Moreover, outside of a spherical body, the spacetime must be Schwarzchild spacetime.  There is also an extension of the theorem to spherically symmetric solutions of the electro--vacuum field equations, in that the spacetime must be Reissner--Nordstr\"{o}m.

There is a version of the Birkhoff package for nonzero cosmological constant as well, see \cite{SW09}.

Israel's Theorem (after Werner Israel) is a related uniqueness result which shows that the only spacetimes which are stationary and asymptotically flat are Schwarzchild, Kerr, and Reisnner--Nordstr\"{o}m.  A generalization of Israel's Theorem is known as the ``no-hair theorem" and posits further that all stationary black hole solutions to the Einstein equations with electromagnetic stress energy tensor are uniquely determined by three quantities: mass, angular momentum, and electric charge.

\end{document}